\PassOptionsToPackage{square,comma,numbers,sort&compress}{natbib}
\documentclass{article}

\usepackage[final]{neurips_2021}

\usepackage[utf8]{inputenc} 
\usepackage[T1]{fontenc}    
\usepackage{url}            
\usepackage{natbib} 
\usepackage{booktabs}       
\usepackage{amsfonts}       
\usepackage{nicefrac}       
\usepackage{microtype}      
\usepackage{xcolor}         
\usepackage[frozencache,cachedir=.]{minted}
\usepackage[ruled,vlined,shortend, linesnumbered]{algorithm2e} 
\usepackage[]{algorithmic} 
\usepackage{tikz,tikzpeople}
\usepackage{amsmath}
\usepackage{todonotes}
\usepackage{wrapfig}
\usetikzlibrary{shapes.geometric}
\usepackage{graphicx}
\usepackage[inline]{enumitem}
\usepackage{pgfplots}
\pgfplotsset{compat=1.5}
\usepgfplotslibrary{groupplots}

\usepackage[hidelinks]{hyperref}       
\definecolor{purple}{rgb}{0.7,0,1.0}

\newcommand{\SpA}{{\textsc{SpliceOut}}\xspace}

\title{\SpA: A Simple and Efficient Audio Augmentation Method}

\author{%
  Arjit Jain \\
  Indian Institute of Technology Bombay\\
  \texttt{arjit@cse.iitb.ac.in} \\
   \And
   Pranay Reddy Samala \\
   Indian Institute of Technology Bombay\ \\
   \texttt{pranayr@cse.iitb.ac.in} \\
   \AND
   Deepak Mittal \\
   Verisk Analytics \\
   \texttt{deepak.mittal@verisk.com} \\
   \And
   Preethi Jyothi \\
   Indian Institute of Technology Bombay\ \\
   \texttt{pjyothi@cse.iitb.ac.in} \\
   \And
   Maneesh Singh \\
   Verisk Analytics \\
   \texttt{maneesh.singh@verisk.com} \\
}

\begin{document}
\maketitle

\begin{abstract}
Time masking has become a \textit{de facto} augmentation technique for speech and audio tasks, including automatic speech recognition (ASR) and audio classification, most notably as a part of SpecAugment. 
In this work, we propose \SpA, a simple modification to time masking which makes it computationally more  efficient. \SpA performs comparably to (and sometimes outperforms) SpecAugment on a wide variety of speech and audio tasks, including ASR for seven different languages using varying amounts of training data, as well as on speech translation, sound and music classification, thus establishing itself as a broadly applicable audio augmentation method. \SpA also provides additional gains when used in conjunction with other augmentation techniques. Apart from the fully-supervised setting, we also demonstrate that \SpA can complement unsupervised representation learning with performance gains in the semi-supervised and self-supervised settings. 
\looseness=-1
\end{abstract}

\section{Introduction}
Data augmentation has become an integral part of the modern machine learning pipeline. It involves applying label-preserving transformations to training data that increases diversity in the training samples, thus acting as a regularizer to prevent overfitting. Data augmentation has also emerged as an important technique in recent self-supervised and semi-supervised learning algorithms to help learn useful representations of the data~\citep{xie2020selftraining,grill2020bootstrap,chen2020exploring,zbontar2021barlow,caron2020unsupervised,he2020momentum,chen2020simple}. Augmentation techniques demonstrated a lot of initial success in image classification tasks~\citep{perez2017effectiveness} and have been adopted for text, speech and audio tasks as well~\citep{wei-zou-2019-eda,kobayashi-2018-contextual,hou-etal-2018-sequence}. For speech and audio-based applications, augmentations like vocal tract length transformations~\citep{jaitly2013vocal}, speed perturbations~\citep{speedpertubation} and adding environment noise were early techniques that were found to be effective. However, these techniques typically incur additional computational costs and sometimes require additional data (e.g. in noise-based augmentations).
\looseness=-1

SpecAugment~\citep{specaugment} offered a simple alternative of directly transforming an audio spectrogram (i.e. a visual representation of the audio signal) by randomly warping across the time axis or masking consecutive chunks of time (i.e. \emph{time masking}) or frequency channels (i.e. \emph{frequency masking}) in the spectrogram. It was shown to be very effective for ASR, and has since been used for audio classification tasks as well~\citep{panns,urban,deepspectrumlite}. Investigations into new augmentation techniques that work broadly across different audio and speech tasks have been fairly limited. This could be partly attributed to the opaque nature of the audio spectrogram. Unlike image transformations (e.g., translation, shear, brightness, etc.) or text transformations (e.g., synonym replacements, word swapping, etc.) whose effects on a training instance can be easily visualized and understood, transformations on audio spectrograms are harder to interpret, and thus more challenging to define.
\looseness=-1

\begin{figure}[t!]
    \centering
    \includegraphics[width=\textwidth]{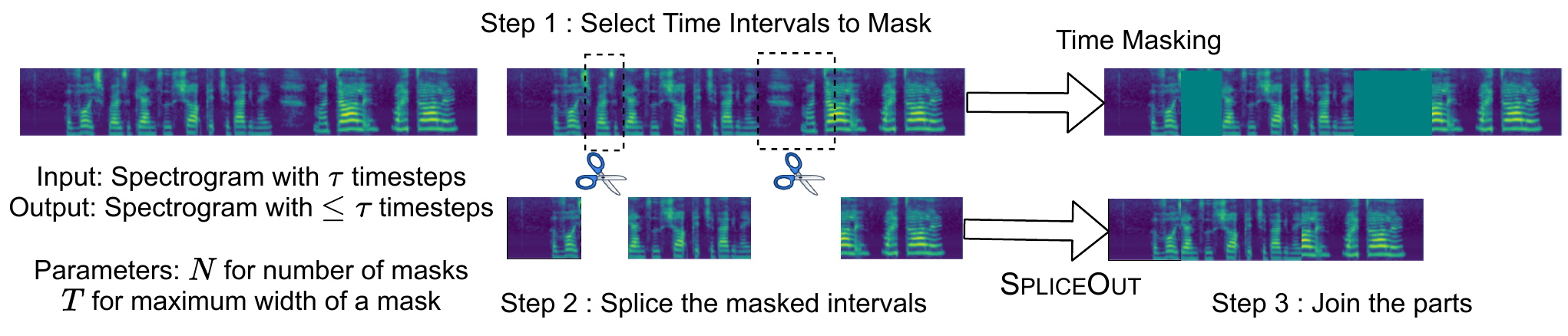}
    \caption{Illustration of \SpA and time masking.} 
    \label{fig:spliceout}
\end{figure}

In this work, we propose \SpA, a simple modification to time masking that makes it more computationally efficient. Instead of masking splices of consecutive time-steps in the audio input, \SpA operates by entirely removing these time slices from the audio input and stitching the remaining parts together. Figure~\ref{fig:spliceout} illustrates the main difference between \SpA and time masking. 
\looseness=-1

While augmentation techniques in prior work have been typically proposed for specific target tasks (e.g. SpecAugment for ASR), we present \SpA as a broadly applicable augmentation technique for most sequence labeling or multi-class classification tasks using speech or audio as input. We substantiate the claim that \SpA has broad applicability by showing that it performs as well as (and at times better than) time masking on many speech and audio tasks including ASR, speech translation, sound and music classification. (\SpA can be applied to both raw waveform and spectrogram-based input formats.) 
We demonstrate the effectiveness of \SpA in both low and high resource training regimes and show ASR results for a number of different languages. Besides fully supervised tasks, we also demonstrate that \SpA can benefit representation learning by showing improvements in semi-supervised and self-supervised settings. Along with establishing that \SpA is useful as a standalone technique, we also show that \SpA can be complementary to time masking. It is noteworthy that in all our experiments \SpA was used as a drop-in replacement for SpecAugment whose hyperparameters were presumably tuned for the task at hand; any performance improvements we derived with \SpA were  without using any separate hyperparameter tuning. 

\section{Related Work}

Masking-based augmentation (and regularization) techniques have been used extensively in machine learning, from masking weights~\citep{germain2015made} and hidden states~\citep{dropout,autodropout,ghiasi2018dropblock,pmlr-v28-wan13} in neural networks, to masking pixels in images~\citep{devries2017improved,Zhong_Zheng_Kang_Li_Yang_2020,8100170,10.1145/344779.344972}, words in text~\citep{devlin2018bert,donahue-etal-2020-enabling,gal2016theoretically}, and time-steps in audio~\citep{specaugment,clar} and time-series data. While masking has been employed as a commonly used augmentation technique, it has also been used in representation learning where the objective involves learning to reconstruct masked out information~\citep{donahue-etal-2020-enabling,10.1145/344779.344972,devlin2018bert}; this reconstruction objective forms the basis of most of the recent semi-supervised and self-supervised learning approaches. Masking-based augmentation techniques can also be used with contrastive losses for representation learning (that have seen a recent resurgence~\citep{zbontar2021barlow,chen2020simple,he2020momentum} where different augmented views of the same data samples can serve as positive examples.

\paragraph*{Audio Augmentations.} Audio augmentation techniques can be broadly classified into the following categories: \begin{enumerate*}
\item \textbf{Warping-based:} Changing the speed of the audio signal using time warping techniques was shown to be effective as an augmentation for ASR~\citep{speedpertubation}.
Both time warping~\citep{specaugment} and frequency warping~\citep{jaitly2013vocal} have been used to augment training examples for ASR.
\item \textbf{Mixing-based:} Mixup~\citep{zhang2018mixup} involves creating new training samples using linear combinations of existing data samples (and their corresponding labels). Mixup was shown to be very effective for audio classification~\citep{panns} and it has also been used successfully for ASR~\citep{mixupasr,meng2021mixspeech}. Related to mixing, reordering shuffled patches of a spectrogram or swapping parts of a spectrogram have also been explored for audio augmentation ~\citep{Carr_2021,specswap}.
\item \textbf{Masking-based:} Time masking and frequency masking~\citep{specaugment} are the most popular audio masking techniques that work by zeroing out randomly chosen blocks of consecutive time steps and frequency channels, respectively. CutMix~\citep{yun2019cutmix} is a recent augmentation technique combining mixing and masking that has been shown to work well for audio classification~\citep{deepspectrumlite}.
\item \textbf{Noise-based:} Augmenting training data with synthesized noisy speech samples~\citep{deepspeech} and simulated reverberant conditions~\citep{ko2017study,kim2017generation} was found to improve ASR on noisy speech and far-field ASR.
\end{enumerate*} 

In addition to the above-mentioned methods, augmentation techniques specific to a target task, such as using text-to-speech synthesis for ASR~\citep{tjandra2017listening,karita2019semi,wang2020improving}, extending masking and mixing to hidden states for audio classification~\citep{specaugmentpp} have also demonstrated benefits.

\SpA shares similarities with frame skipping/dropping methods~\cite{dynamicframeskipping,45555,7472084,6639137} in that both offer computational advantages by ignoring parts of the input. We compare \SpA with these methods in Appendix \ref{app:frameskipping}

\SpA would qualify as a time masking technique, except we delete blocks of consecutive time-steps instead of masking them. In subsequent experimental comparisons, we use \SpA as a drop-in replacement for time masking. \SpA is also complementary to other augmentation techniques as will be demonstrated in Section~\ref{sec:expts}.

\section{\SpA}
\SpA is parameterized by $N$, the number of time intervals to splice, and $T$, the maximum width of a time interval. Given a log-mel spectrogram with $\tau$ time steps, \SpA selects $N$ intervals, each of the form $[t_0, t_0+t)$ where $t \in \mathbb{Z}^{0+}$ is sampled from the uniform distribution $U(0, T)$, and $t_0 \in \mathbb{Z}^{0+}$ is chosen from $[0, \tau - t)$. The input spectrogram is modified by removing all time steps in the union of the selected intervals. NumPy-style pseudocode for both \SpA and Time-masking is outlined below in Algorithms~\ref{alg:spliceout} and~\ref{alg:tm}. 

\begin{minipage}{0.46\linewidth}
\begin{algorithm}[H]
\caption{Pseudocode for \SpA}
\label{alg:spliceout}
\begin{minted}{python}
def SpliceOut(spec, N, T):
  tau = spec.shape[0]
  mask = np.ones(tau, dtype=bool)
  for i in range(N):
    length = randint(T)
    start = randint(tau-length)
    mask[start:start+length] = False
  spec = spec[mask]
  return spec
\end{minted}
\end{algorithm}
\end{minipage}
\hfill
\begin{minipage}{0.46\linewidth}
\begin{algorithm}[H]
\caption{Pseudocode for Time-Masking}
\label{alg:tm}
\begin{minted}{python}
def TimeMasking(spec, N, T):
  tau = spec.shape[0]
  for i in range(N):
    length = randint(T)
    start = randint(tau-length)
    spec[start:start+length] = 0
  return spec
\end{minted}
\end{algorithm}
\end{minipage}
\paragraph{Computational Efficiency.} 

Figure \ref{fig:spliceout} illustrates how \SpA decreases the length of the output spectrogram, as opposed to Time-Masking which results in an output spectrogram of the same length as the original input spectrogram. As we increase the amount of masking, \SpA deletes larger numbers of time intervals from the input data, and the resulting shorter inputs in turn reduce the memory footprint, and time, required for training. Time-Masking, on the other hand, does not gain any computational advantage with respect to the memory or the time required for training, with increased amount of masking.

%
\begin{figure}
\label{fig:maskimprovements}
\centering
\begin{tikzpicture}
\pgfplotsset{
      scale only axis,
      width=0.3\textwidth,height=0.15\textwidth
  }

    



\begin{groupplot}[group style={group size= 2 by 1,ylabels at=edge left, horizontal sep=1.75cm},width=0.36\textwidth,height=0.12\textwidth]
\nextgroupplot[label style={font=\normalsize},
xtick = {2, 4, 8, 16, 32, 64},
xmode = log,
log ticks with fixed point,
tick label style={font=\normalsize},
legend style={at={($(0,0)+(1cm,1cm)$)},legend columns=7,fill=none,draw=black,anchor=center,align=center},
ylabel = {Time per step (in s)},
legend to name=fredself1,
title = Time Improvement,
mark size=1pt]
\addplot[red!70!black,mark=*] coordinates {(2,6.024096386)
    (4,5.988023952)
    (8,6.060606061)
    (16,6.060606061)
    (32,6.060606061)
    (64,6.060606061)
    };
    \addplot[brown,mark=star] coordinates {(2,5.94530321)
    (4,5.820721769)
    (8,5.60758145)
    (16,5.128205128)
    (32,4.308487721)
    (64,3.138731952)
    };
\addlegendentry{Time-Masking};    
\addlegendentry{\SpA};
\coordinate (c1) at (rel axis cs:0,1);

\nextgroupplot[label style={font=\normalsize},
xtick = {2, 4, 8, 16, 32, 64},
xmode = log,
log ticks with fixed point,
tick label style={font=\normalsize},
ylabel = {Memory (in GiB)},
title = Memory Improvements,
mark size=1pt]
\addplot[red!70!black,mark=*] coordinates {(2,12.12207)
    (4,12.12207)
    (8,12.12207)
    (16,12.12207)
    (32,12.12207)
    (64,12.12207)
    };
    \addplot[brown,mark=star] coordinates {(2,12.026367)
    (4,11.764648)
    (8,11.581055)
    (16,11.084961)
    (32,9.9345703)
    (64,8.37207)
    };     
\end{groupplot}

\end{tikzpicture}
{\pgfplotslegendfromname{fredself1}};
\caption{Comparison of running time and memory requirements during training using Time-Masking and \SpA augmentations, with varying number of masks.}
\label{fig:efficiency}
\end{figure}
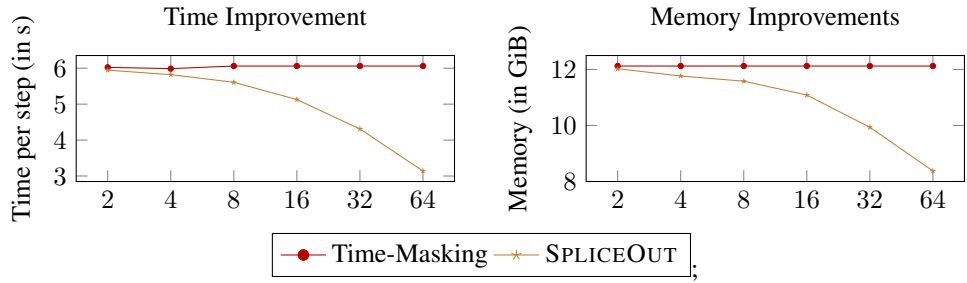

We empirically verify this for an ASR task on the LibriSpeech $100$-hr benchmark. (Details of the experimental setup and the model configuration are described in Section \ref{sec:libri}.) 
Figure \ref{fig:efficiency} plots the time taken per training step, and the memory required for training, for both Time-Masking and \SpA, for different values of $N$, while keeping the maximum width $T$ constant at $40$ time steps. As expected, we observe that \SpA is indeed more time-efficient and memory-efficient than Time-Masking. At an extreme level of masking i.e., $N=64$, \SpA achieves almost $2\text{x}$ speedup in training, and requires $33\%$ less memory. For a moderate masking level, $N=8$, \SpA achieves $8\%$ speedup in training, and requires $\sim 5\%$ less memory. For a fixed masking budget, we see that \SpA is more computationally efficient than Time-Masking, with larger gains as we increase the amount of masking. (ASR performance on Librispeech $100$-hr with varying number of masks is given in Section~\ref{sec:libri}.) 

\section{Experiments}
\label{sec:expts}

We demonstrate the effectiveness of \SpA on a wide variety of speech and audio tasks spanning different model architectures,  learning paradigms and toolkits. Table~\ref{tab:summary} summarizes the main tasks and datasets used in our experiments. We start with a well-known benchmark for English ASR, Librispeech~\citep{librispeech}. We present low-resource ASR results for many languages from the CommonVoice corpus~\citep{commonvoice} and show significant improvements in performance using \SpA. We also show performance improvements using \SpA on a speech translation task  Libri-Trans~\citep{libritrans}, translating English speech into French text. Apart from the sequence labeling tasks, we also evaluate \SpA using two popular sound classification tasks, ESC-50 and UrbanSound8K, and a music genre classification task. Finally, we show how \SpA helps learn better speech representations when trained using semi-supervised and self-supervised objectives. 

For significance analyses on ASR, we use Bootstrap Confidence Intervals (BCI)~\citep{bootstrap}, a widely adopted method for ASR performance evaluation, and for classification, we use the standard deviation obtained from cross-validation on predefined folds. Additional details about the significance and confidence analyses can be found in Appendix. 



 Machine specifications and complete implementation details can be found in Appendix A.1. Appendix A.2 describes the datasets and frameworks we use, along with their respective licenses. Appendix A.3 explores the use of TM and \SpA with Sample Adaptive Policies~\citep{sapaugment} that help learn policies to automatically combine augmentation methods. Appendix A.4 shows how \SpA can be used to learn with noisy data, Appendix A.5 compares TM and \SpA as data augmentations to fine-tune a pre-trained wav2vec2 model using raw waveforms from TIMIT as input, and Appendix A.6 uses \SpA with raw waveforms for representation learning. Appendix A.7 supplements results from Table~\ref{tab:commonvoice} by extending to more languages from the CommonVoice corpus. 

\begin{table}
    \caption{A collated summary of the different tasks used to evaluate \SpA, along with the corresponding input representations, types of tasks, models and datasets.}
    \centering
    \begin{tabular}{llllll}
    \toprule
    Audio as & Task & Format & Type & Model & Datasets\\
    \midrule
    Speech & Recognition & Spectrogram & High Resource & Conformer & LibriSpeech-100\\
    &&&& Transformer & \\
    &&& Low Resource & Conformer & CommonVoice\\
    && Waveform & Fine-Tuning & Wav2Vec2 & TIMIT\\
    & Translation & Spectrogram & High Resource & & LibriTrans (En-Fr)\\
    & Classification & Spectrogram & Semi-Supervised & Resnet18 & SpeechCommands\\
    &&& Self-Supervised && \\
    && Waveform & Semi-Supervised & Resnet18 & \\
    &&& Self-Supervised &  & \\
    \midrule
    Sound & Classification & Spectrogram & Fine-Tuning & CNN10 & ESC-50, US8K\\
    &&& Self-Supervised & Resnet-18 & FSDNoisy18K\\
    &&& Pseudo-Labeling & DenseNet & \\
    \midrule
    Music & Classification & Spectrogram & Fine-Tuning & CNN14 & GTZAN \\
    \bottomrule
    \end{tabular}
    \label{tab:summary}
\end{table}


\subsection{Speech-based Sequence Labeling Tasks}

\subsubsection{ASR: LibriSpeech}
\label{sec:libri}

Librispeech is a widely used ASR benchmark consisting of English read speech. We use the clean $100$-hr subset for training and evaluate on the standard dev (clean/other) and test (clean/other) sets. The ``other" evaluation datasets comprise speech samples that are more acoustically challenging. Our base model is the large variant of the Conformer model~\citep{conformer}, Conformer(L), which is a state-of-the-art network for ASR and is implemented using the ESPnet toolkit~\citep{espnet}. We use $83$-dimensional log mel-filterbank + pitch features as input. Each model is trained for $60$ epochs, and model averaging is performed on weights from $5$ epochs with the best performance on the validation set. Speed perturbation, with speeds $0.9, 1.0, 1.1$, is applied to the training data. Beam search is used during decoding, without invoking any external language model.


\begin{table}
\centering
    \caption{WERs on LibriSpeech test sets, using TM, FM and \SpA (SO), with $N=2$.}
    \begin{tabular}{lrr}
    \toprule
     Augmentation & test-clean & test-other\\
    \midrule
    TM & $7.6{\scriptstyle \pm 0.19}$ & $21.8{\scriptstyle \pm 0.31}$\\
    SO & $\mathbf{7.5}{\scriptstyle \pm 0.18}$ & $\mathbf{21.4}{\scriptstyle \pm 0.31}$\\
    \midrule
    FM + TM & $\mathbf{7.2}{\scriptstyle \pm 0.17}$ & $18.3{\scriptstyle \pm 0.28}$\\
    FM + SO & $\mathbf{7.2}{\scriptstyle \pm 0.18}$ & $\mathbf{18.2}{\scriptstyle \pm 0.29}$\\
    \midrule
    TW + FM + TM~\citep{specaugment,espnet} & $\mathbf{7.0}{\scriptstyle \pm 0.16}$ & $18.1{\scriptstyle \pm 0.28}$\\
    TW + FM + SO & $7.1{\scriptstyle \pm 0.17}$ & $17.9{\scriptstyle \pm 0.29}$\\
    TW + FM + TM + SO & $7.1{\scriptstyle \pm 0.18}$  & $\mathbf{17.7}{\scriptstyle \pm 0.29}$\\
    \bottomrule
    \end{tabular}
    \label{tab:libriaugs}
\end{table}

Table~\ref{tab:libriaugs} shows the word error rates (WERs) using \SpA as a replacement for Time-Masking (TM), both with and without the presence of other augmentations like time warping (TW) and frequency masking (FM). The time warp parameter for TW is set to $5$. We use $2$ masks of maximum width $30$ for FM. For TM and \SpA, $N=2$ and $T=40$. We observe that \SpA either performs comparably or a bit better when used as a replacement for TM. Interestingly, using \SpA in conjunction with SpecAugment (TW, FM, TM) yields further improvements in performance on test-other. Table \ref{tab:librimasks} shows how WERs change with increasing $N$ for TM and \SpA. For a fixed $N$, we observe that \SpA almost consistently outperforms TM. (With larger values of $N \ge 16$, the performance starts to degrade.)

\begin{table}
    \centering
    \caption{Effect of increasing the number of masks, $N$, in Time-Masking and \SpA (SO) augmentations, on WERs of LibriSpeech test sets.}
    \begin{tabular}{rlrrrr}
    \toprule
    N & Method & test-clean & test-other\\
    \midrule
    $2$ & TM & $7.6 {\scriptstyle \pm 0.19}$ & $21.8{\scriptstyle \pm 0.31}$\\
    & SO & $\mathbf{7.5}{\scriptstyle \pm 0.18}$ &  $\mathbf{21.4}{\scriptstyle \pm 0.31}$\\
    \midrule
    $4$ & TM  & $7.3{\scriptstyle \pm 0.18}$  & $20.4{\scriptstyle \pm 0.30}$\\
    & SO & $\mathbf{7.0}{\scriptstyle \pm 0.16}$ & $\mathbf{20.3}{\scriptstyle \pm 0.31}$\\
    \midrule
    $8$ & TM & $\mathbf{6.8}{\scriptstyle \pm 0.17}$  & $19.2{\scriptstyle \pm 0.29}$\\
    & SO  & $\mathbf{6.8}{\scriptstyle \pm 0.18}$ & $\mathbf{19.0}{\scriptstyle \pm 0.30}$\\
    \bottomrule
    \end{tabular}
    \label{tab:librimasks}
\end{table}

\paragraph*{Semantic Mask.} While SpecAugment randomly masks time intervals in a spectrogram, Semantic-Mask~\citep{semanticmask} proposes masking out regions which correspond to a word or word-piece in the transcription, thus encouraging the decoder to learn a better internal language model. Similar to how we modify TM to implement \SpA, we modify Semantic-Mask to implement Semantic-Splice wherein we splice, or delete, certain intervals which correspond to a word or word-piece. We compare Semantic-Mask and Semantic-Splice on the LibriSpeech $100$-hr dataset. We follow the experimental setup described in ~\citep{semanticmask}, i.e. we use the Montreal Forced Aligner to compute forced alignments between the acoustic features and transcriptions. $15\%$ of the tokens are masked or spliced out in Semantic-Mask and Semantic-Splice, respectively. A transformer model is used as the base architecture, with convolutional layers for encoding, instead of positional encodings, similar to~\citep{semanticmask}. Speed pertubation, with speeds $0.9, 1.0, 1.1$ is used, along with SpecAugment. Models for both Semantic-Mask and Semantic-Splice were trained for $120$ epochs, and the model with the best validation performance was used for testing. Table \ref{tab:semantic} reports WERs for both methods showing that Semantic-Splice consistently outperforms Semantic-Mask on both the test sets. 

\begin{table}[h]
    \centering
    \caption{WERs on LibriSpeech test sets comparing Semantic-Mask and Semantic-Splice.}
    \begin{tabular}{lrr}
    \toprule
     Method & test-clean & test-other\\
    \midrule
    Semantic-Mask~\cite{semanticmask} & $9.2$ & $22$\\
    Semantic-Splice (Ours) & $\mathbf{8.8}$ & $\mathbf{21.5}$\\
    \bottomrule
    \end{tabular}
    \label{tab:semantic}
\end{table}


\subsubsection{ASR for Multiple Languages: CommonVoice}
\label{sec:commonvoice}
The CommonVoice corpus~\citep{commonvoice} is a multilingual corpus of read speech in 38 languages. We train ASR models for six different languages spanning training sizes ranging from 10 hours to 96 hours: Kyrgyz, Swedish, Tatar, Turkish, Ukranian and Welsh. We use a conformer architecture adapted for ESPnet\citep{espnetconformer} (more details are in Appendix A.1). The same model architecture and hyperparameter settings are used for all languages. We train the model for $50$ epochs with $83$-dimensional features ($80$ log-mel filterbank and $3$ pitch) and byte pair encoding (BPE)-based encoded transcripts (with a vocabulary of size 150. Speed perturbation is applied, with speeds $0.9$, $1.0$ and $1.1$, during training. Evaluation is done by averaging the $10$ models with the best validation accuracy. Beam search decoding without external language model is used during inference. Parameter values for TW, FM, TM are the same as in Section~\ref{sec:libri}.

Table~\ref{tab:commonvoice} shows WERs for all six languages. Our experiments demonstrate that \SpA consistently outperforms TM, and  obtains statistically significant WER reductions on four low-resource languages, Kyrgyz, Swedish, Tatar and Ukrainian (at $p<0.05$ using the MAPSSWE test preferred for ASR evaluations~\citep{gillick1989some}). 

\begin{table}[h!]
    \centering
    \caption{Evaluation of TM and \SpA, when used with TW and FM, using test WERs across multiple different languages, including several low-resource settings.}
    \begin{tabular}{l|rrrrr|r}
    \toprule
     Augmentation & Swedish  & Turkish  &  Kyrgyz & Ukrainian & Tatar & Welsh\\
    TW + FM & 10 hrs & 22 hrs & 22 hrs & 25 hrs & 28 hrs & 96 hrs\\ 
    \midrule
     + TM~\citep{specaugment,espnet} & $33.2$ &  $6.7$  & $37.3$ & $14.6$ & $36.3$ & $15.4$\\
     + \SpA & $\mathbf{32.1}$ & $\mathbf{6.5}$ & $\mathbf{36.2}$ & $\mathbf{14.0}$ & $\mathbf{35.5}$ & 
      $\mathbf{14.8}$\\
    \bottomrule
    \end{tabular}
    \label{tab:commonvoice}
\end{table}

\subsubsection{Speech Translation: Libri-Trans}
Libri-Trans is a speech translation (ST) benchmark with approximately $240$h of English read speech (from Librispeech) aligned with French text ~\citep{libritrans}. The base model for ST consists of an ASR encoder and a machine translation (MT) decoder. We initialized the encoder with a SpecAugment-pretrained ASR Transformer and the decoder with a pretrained MT Transformer. Both the models employ BPE units with a vocabulary of size 1K and use joint source and target vocabularies. Speed perturbation, with speeds 0.9, 1.0 and 1.1, was used for ST training.  Finetuning is performed for $50$ epochs, and model averaging is performed on weights from the $5$ epochs with best validation accuracy for evaluation. The ST experiments were conducted using the ESPNet-ST framework~\citep{espnetst}. Evaluations are performed on the predefined dev and test splits mentioned in Libri-Trans. Table~\ref{tab:bleu} shows marginal improvements in BLEU scores on the dev and test sets with using \SpA as opposed to TM. 
 

\begin{table}[]
    \centering
    \caption{Evaluating TM and \SpA using development and test set BLEU scores for the Libri-Trans task. Higher is better.}
    \begin{tabular}{lrr}
    \toprule
    Augmentation & Dev BLEU & Test BLEU\\
    \midrule
         TW + FM + TM~\citep{specaugment,espnetst} & $18.43$ & $17.18$\\
         TW + FM + \SpA & $\mathbf{18.57}$ & $17.15$\\
         TW + FM + TM + \SpA & $18.42$ & $\mathbf{17.35}$\\
    \bottomrule
    \end{tabular}
    \label{tab:bleu}
\end{table}

\subsection{Audio Classification Tasks}

\subsubsection{Sound Classification: ESC-50 and UrbanSound8K}
\label{sec:soundclass}

We compare \SpA with Time-Masking on audio classification Tasks. Environmental Sound Classification (ESC-50) ~\citep{piczak2015dataset} and UrbanSound8K ~\citep{urban} are two well-established benchmarks in sound classification containing $50$ and $10$ sound classes, respectively. State-of-the-art techniques on these datasets rely on transfer learning, where a network pretrained on a large audio classification dataset like AudioSet ~\citep{audioset} is further fine-tuned on labeled data. We use the CNN10 architecture~\citep{panns} that is pretrained on AudioSet and takes mel-spectrograms as input. The experimental setup is similar to~\citep{urban}.
\looseness=-1

Standard evaluation protocols defined for these benchmarks are followed: For ESC-50, cross-validation is performed over the $5$ pre-defined folds, and for UrbanSound8k, cross-validation is performed over the $10$ pre-defined folds. (More implementation details can be found in Appendix A.1.)
Along with TM, and/or \SpA, Mixup (MX) and Frequency Masking (FM) are used for data augmentation. We use $2$ masks for FM, with a maximum frequency width of $8$. The $\alpha$ parameter for mixup is set to $1$. For TM and \SpA, $N=2$ and $T=24$. 

\begin{table}
    \caption{Evaluating TM and \SpA on two sound classification tasks, with the standard augmentation combinations~\citep{urban,panns}. Higher is better.}
    \centering
    \begin{tabular}{l|rrr}
    \toprule
         Augmentation & Accuracy & F1$_\text{micro}$ & mAP \\
    \midrule
    \multicolumn{4}{c}{Dataset: ESC-50}\\
    \midrule
     MX + FM + TM & $90.40{\scriptstyle \pm 0.02}$ & $89.42{\scriptstyle \pm 0.02}$  & $94.98{\scriptstyle \pm 0.01}$\\
     MX + FM + SO & $90.95{\scriptstyle \pm 0.02}$ & $89.96{\scriptstyle \pm 0.02}$ & $95.17{\scriptstyle \pm 0.01}$\\
    \midrule
    \multicolumn{4}{c}{Dataset: UrbanSound8k}\\
    \midrule
    MX + FM + TM & $86.39{\scriptstyle \pm 0.04}$ & $\mathbf{86.32}{\scriptstyle \pm 0.04}$ & $\mathbf{93.04}{\scriptstyle \pm 0.03}$\\
    MX + FM + SO & $\mathbf{86.67}{\scriptstyle \pm 0.04}$ & $\mathbf{86.31}{\scriptstyle \pm 0.04}$ & $\mathbf{93.04}{\scriptstyle \pm 0.03}$\\
    \bottomrule
    \end{tabular}
    \label{tab:soundclass}
\end{table}
\begin{table}
    \centering
    \caption{Evaluating TM and \SpA, with and without FM, on GTZAN Music Genre Classification. \SpA is complementary to TM. Higher is better.}
    \begin{tabular}{l|r}
    \toprule
         Augmentation & Accuracy \\
    \midrule
    MX + TM & $90.7{\scriptstyle \pm 0.03}$\\
    MX + SO & $90.7{\scriptstyle \pm 0.03}$\\
    \midrule
    MX + FM + TM~\citep{panns} & $91.3{\scriptstyle \pm 0.03}$\\
    MX + FM + SO & $\mathbf{91.4}{\scriptstyle \pm 0.03}$\\
    \midrule
    MX + FM + TM + SO & $\mathbf{92.8}{\scriptstyle \pm 0.03}$\\
    \bottomrule
    \end{tabular}
    \label{tab:music}
\end{table}


Table~\ref{tab:soundclass} shows that \SpA consistently improves performance on all three metrics for ESC-50, Accuracy, F1$_{\text{micro}}$ and mean AP compared to TM, and performs as well as TM on UrbanSound8K.

\subsubsection{Music Classification: GTZAN}

We also consider the task of music genre classification using the GTZAN dataset~\citep{gtzan} consisting of $10$ genre labels. Similar to Section \ref{sec:soundclass}, we follow the experimental setup in~\citep{panns} and use the CNN14 architecture, pretrained on AudioSet, which is then further fine-tuned on data from GTZAN. The model is trained for $1000$ iterations with a batch size of $16$. For evaluation, $10$-fold cross validation is performed. Implementation details can be found in Appendix A.1. The parameters used for FM and MX are the same as in Section~\ref{sec:soundclass}; for TM and \SpA, $N=2$ and $T=64$. Table~\ref{tab:music} shows that using \SpA in place of TM results in comparable performance, while using \SpA in conjunction with TM gives a further boost to test set accuracy.



\subsection{Representation Learning: CLAR}

Contrastive Learning of Audio Representation (CLAR)~\citep{clar} builds on SimCLR~\citep{chen2020simple} to provide a framework for semi-supervised and self-supervised representation learning for audio data using multiple augmentations. Similar to SimCLR, CLAR learns representations by maximizing the similarity between augmented views of the same data samples, and minimizing the similarity with ``negative" samples, taken to be all other samples in the mini-batch. The Normalized Temperature-scaled Cross Entropy loss for a positive pair of examples indexed by $(i, j)$ is given by:

$$
\mathcal{L}_{CL}=-\sum_{i, j}^{N} \log \frac{\exp \left(\operatorname{sim}\left(\mathbf{z}_{i}, \mathbf{z}_{j}\right) / \tau\right)}{\sum_{k=1}^{2 N} \mathbf{1}_{[k \neq i]} \exp \left(\operatorname{sim}\left(\mathbf{z}_{i}, \mathbf{z}_{k}\right) / \tau\right)}
$$
where $\tau$ is the temperature, $N$ is the mini-batch size and $\mathbf{z}$ represents the projected representations. This loss is computed across all positive pairs, both $(i, j)$ and $(j, i)$, in the mini-batch. In the semi-supervised setting, CLAR uses both $\mathcal{L}_{CL}$, and the standard cross entropy loss, $\mathcal{L}_{CE}$ to train the model. For samples without any labels, $\mathcal{L}_{CE}$ is set to $0$. 

ResNet18 is the base encoder model with the output dimension set to $512$. The projection head used to generate $\mathbf{z}$ is implemented as three fully-connected layers with ReLU activations. To evaluate the learned representations, a linear classifier is trained on the frozen features, as has been done in prior work~\citep{grill2020bootstrap,chen2020exploring,zbontar2021barlow,chen2020simple}. ~\citep{clar} recommends using Fade In/Out (FD) and Time-Masking (TM) to generate augmented views of the same audio sample. In FD, the intensity is varied on the boundaries of the audio signal. The degree and the size of fade are selected stochastically for each sample. We use the same experimental setup as CLAR, i.e. we perform our experiments on the Speech Commands-10 dataset consisting of speech samples corresponding to $10$ isolated-word commands. The batch size is set to $512$, and global batch normalization is used.%
\footnote{We found the SGD optimizer to be more stable during training, compared to the Adam optimizer used in CLAR. We use Layer-wise Adaptive Rate Scheduling (LARS), with a learning rate of $1$, weight decay $10^{-4}$, with linear warm-up for the first $10$ epochs; the learning rate is decayed with a cosine schedule without restarts.}
More implementation details and experiments with semi-supervised and self-supervised methods using the raw speech waveform can be found in Appendix A.1, and A.6 respectively.
\looseness=-1

\begin{table}[h]
    \centering
    \caption{Comparing classification accuracies using \SpA and TM in semi-supervised (with different amounts of labeled data) and self-supervised settings on the Speech Commands Dataset. Higher is better.}
    \begin{tabular}{ll|rrrr}
    \toprule
         & & \multicolumn{4}{c}{Labeled Data Percentage} \\
         Type & Method & $100\%$ & $20\%$ & $10\%$ & $1\%$\\
         \midrule
         Supervised&Cross Entropy& $94.9$ & $86.4$ & $68.4$ & $28.6$\\
         \midrule
         Semi-Supervised&SupCon & $96.0$ & $87.9$ & $82.1$ & $26.6$\\
         &CLAR (FD + TM)~\citep{clar,specaugment} & $97.2$ & $94.7$ & $91.7$ & $\mathbf{72.8}$\\
         &CLAR (FD + \SpA) & $\mathbf{97.4}$ & $\mathbf{95.6}$ & $\mathbf{92.6}$ & $71.2$\\
         \midrule
         Self-Supervised & SimCLR (FD + TM)~\citep{clar,specaugment} & \multicolumn{4}{c}{$89.0$}\\
         & SimCLR (FD + \SpA) & \multicolumn{4}{c}{$88.9$}\\
    \bottomrule
    \end{tabular}
    \label{tab:clar}
\end{table}

Table~\ref{tab:clar} compares TM with \SpA when used in conjunction with FD for both semi-supervised and self-supervised representation learning. The results for Supervised (Cross Entropy) and the baseline system SupCon~\citep{supcon} are reproduced from~\citep{clar}. In 3 out of 4 labeled data settings, \SpA yields performance improvements compared to TM. (Accuracies on the purely self-supervised task i.e. using zero amount of labeled data, are comparable using TM and \SpA.)
\looseness=-1



\section{Discussion}

We present a comprehensive empirical evaluation of \SpA as a generic audio augmentation technique. We show that it works with: \begin{enumerate*}
\item Different types of audio signals (speech in different languages, sounds, music)
\item Different tasks (ASR, speech translation, classification)
\item Different model architectures and optimization functions (encoder-decoder conformer models, CNN10/CNN14 convolutional networks)
\item Different learning paradigms (fully supervised, semi-supervised, self-supervised), and
\item Different toolkits (ESPnet, speechbrain)
\end{enumerate*}. We emphasize that all our results were obtained \emph{without any} hyperparameter tuning. We used \SpA as a drop-in replacement for SpecAugment in the existing implementations. Hyperparameters of the latter were presumably tuned to work well for the specific target tasks. Our results show that, even without any tuning, \SpA is at least as good as (and sometimes better than) Time-Masking.





\paragraph*{Properties of \SpA.}
It has been argued that a data augmentation technique (as opposed to a regularization technique) should result in data that is ``consistent with observations that may be seen by the model''~\citep{wang2020improving}. Motivated by this, in addition to the empirical evaluation of \SpA in terms of accuracy in various tasks, we briefly explore its effects in terms of statistical and perceptual distortion, vis-a-vis Time-Masking.

Firstly, we consider simple time-averaged statistics -- specifically, mean and variance -- of the modified spectrograms.
This is motivated in part by normalization techniques like BatchNorm~\citep{batchnorm}, and by work on auditory perception that provides evidence that the human auditory system uses time-averaged statistics for input summarization~\citep{summary1,summary2}.
We provide an empirical comparison of mean and variance of the spectrogram obtained after applying TM or \SpA augmentations with that of the original signal. In our comparisons, we also include an additional variant, TM (Mean), which explicitly corrects for the mean by mean imputation i.e., setting the masked parts to the mean of the input spectrogram instead of zero as in TM (Zero). As Figure~3 shows, \SpA maintains mean and variance better than TM (Zero); it matches TM (Mean), and provides a better match for variance.

\begin{figure}[h]
\includegraphics[width=\textwidth]{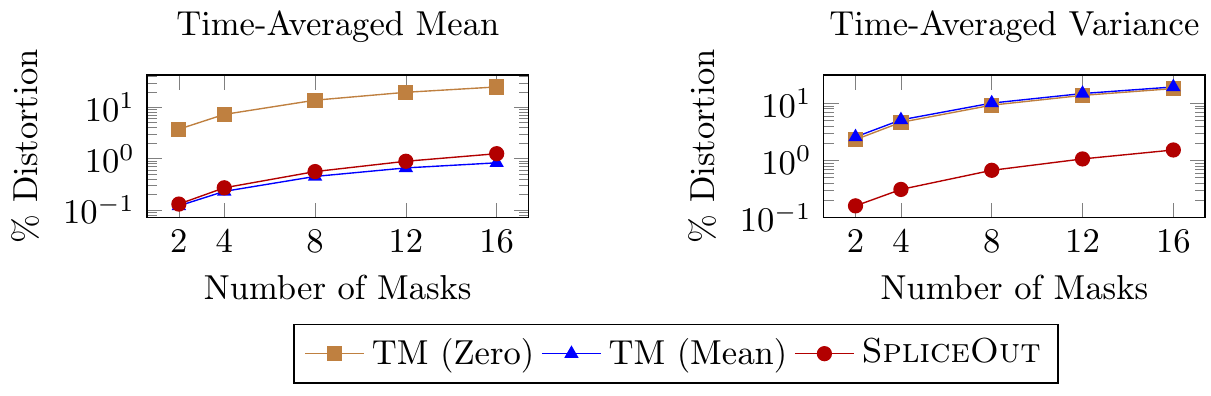}
\label{fig:summstats}
\caption{Comparison of \% distortion in the Time-Averaged Statistics of different augmentation methods, compared to the unaltered input, with varying number of masks.}
\end{figure}

To evaluate perceptual distortion, we use two objective intelligibility and quality measures: Perceptual Evaluation of Speech Quality (PESQ)~\citep{pesq} is one of the most widely used metrics that correlates with mean opinion scores from human evaluations of speech signals. Speech-to-Reverbaration Modulation energy Ratio (SRMR)~\citep{falk2010non} is a more recent metric that was proposed in the context of dereverberated speech. Table~\ref{tab:perceptual} shows PESQ (using both narrowband 8kHz and wideband 16kHz versions) and SRMR values computed for waveforms reconstructed from $100$ augmentated spectrograms each for $100$ random samples from Librispeech using \SpA and TM (zero and mean imputation). \SpA yields consistently higher PESQ scores compared to both TM approaches (and performs at par on SRMR). Thus, in both the metrics explored, \SpA better approximates real-life data, and better fits the notion of a data augmentation technique compared to TM, without incurring any efficiency penalties.

\begin{table}[hbt]
    \centering
    \caption{Perceptual speech metrics, both absolute and relative, comparing the quality of speech modified by TM (Zero), TM(Mean), and \SpA transformations. Higher is better.}
    \begin{tabular}{lrrr}
    \toprule
    & Absolute & \multicolumn{2}{c}{Relative}\\
    \cmidrule{2-4}

    Augmentation & SRMR & Wide-Band PESQ & Narrow-Band PESQ \\
    \midrule
     TM (Zero) & $\mathbf{9.24}$ & $3.07$ & $3.35$  \\
     TM (Mean) & $9.14$ & $3.05$ & $3.46$\\
     \SpA & $\mathbf{9.24}$ & $\mathbf{3.33}$ & $\mathbf{3.59}$ \\
     \bottomrule
    \end{tabular}
    \label{tab:perceptual}
\end{table}




There are a few limitations of \SpA that we aim to address as part of future work. While \SpA is just a simple modification to Time-Masking, it would be hard to design a counterpart of \SpA for frequency masking given the fixed number of frequency channels. Another challenge is to use \SpA with large batch sizes; this might require more padding which, in turn, would counteract the speedup obtained using \SpA. Extending \SpA to hidden states would not be as seamless as extending masking to hidden states~\cite{specaugmentpp}. Splicing out hidden states in networks that optimize the CTC loss~\cite{ctc} could result in input sequences that are shorter than the output sequences, thus violating the CTC prerequisites. 

\section{Conclusions}
\label{sec:conclusions}
We propose \SpA, a new audio augmentation technique that is a simple modification to time masking and that works well for a variety of audio and speech classification tasks. \SpA was shown to significantly outperform time masking in low-resource ASR tasks across many languages. \SpA also offers potential efficiency gains by tuning the number of masks applied to the audio input, unlike time masking where the computational effort is invariant to the amount of masking. We also present analyses to support the claim that \SpA is better at approximating valid speech samples compared to time masking and hence is a better motivated data augmentation technique.

\newpage
\appendix
\section{Appendix}
\subsection{Experimental Details}
\label{sec:implementationdetails}

\paragraph{Machine Configuration. } 
All experiments were conducted on an HPC cluster with $64$ GPU nodes, each containing an NVIDIA P100 GPU with an Intel Broadwell $2.1$Ghz CPU and $64$ GB RAM.
Environment details, including library and package versions, can be found in the code we have included as part of supplementary material.

\subsubsection{ASR: LibriSpeech}
\label{sec:appasr}
We use the ESPnet framework~\citep{espnet} for implementing end-to-end automatic speech recognition with a Pytorch backend\footnote{https://github.com/espnet/espnet/tree/master/egs/librispeech/asr1}. The base conformer model consists of $12$ encoder layers, and $6$ decoder layers, with $0.1$ dropout rate. The encoder and decoder layers contain $2048$ hidden units. The input layer is a $2$D convolutional layer with a kernel size of $31$. Macaron implementation is used as described in \citep{conformer,espnetconformer}. For attention, eight $512$-dimensional self-attention heads are used with relative positional encodings, swish activation and $0$ dropout. The NoAM optimizer, with no patience and with an initial learning rate of $10$, is used with $25000$ warmup steps. For training, a batch size of $32$ is used, with gradient accumulation for $8$ steps and gradients are clipped to $5$. For time warping, the maximum time warp window is set to $5$, for frequency masking, the number of masks $N=2$ and the maximum frequency width is set to $30$. Unless specified otherwise, for Time-Masking and \SpA, $N=2$ and $T=40$. We use $83$-dimensional log mel-filterbank + pitch features as input. Each model is trained for $60$ epochs, and model averaging is performed on weights from $5$ epochs with the best performance on the validation set. Speed perturbation, with speeds $0.9, 1.0, 1.1$, is applied to the training data. Beam search is used during decoding with beam width $60$, without any penalties or length restrictions, and without invoking any external language model. 

\paragraph{Semantic Mask.} We use the official open-sourced ESPnet-based  implementation\footnote{https://github.com/MarkWuNLP/SemanticMask}, and follow the experimental setup described in ~\citep{semanticmask}, i.e., we use the Montreal Forced Aligner\footnote{https://github.com/MontrealCorpusTools/
Montreal-Forced-Aligner} to compute alignments. $15\%$ of the tokens are masked or spliced out in Semantic-Mask and Semantic-Splice, respectively. A transformer model with $12$ encoder layers, $6$ decoders, and an attention vector of size $512$ with $8$ heads, containing $75M$ parameters overall, is used as the base architecture, with convolutional layers for encoding, instead of positional encodings, similar to~\citep{semanticmask}. Models for both Semantic-Mask and Semantic-Splice were trained for $120$ epochs, and the model with the best validation performance was used for testing. SentencePiece was used for tokenization with a vocabulary size of $5000$. The hyper-parameters of SpecAugment, and \SpA were the same as in Section \ref{sec:appasr}. The Adam optimizer was used to update the model with warmup for $25000$ steps and the learning rate decay is proportional to the inverse square root of the step number after the $25000^{\text{th}}$ step.

\subsubsection{ASR for Multiple Languages: CommonVoice}
\label{sec:appcommon}
We use the ESPnet framework~\citep{espnet} for implementing end-to-end automatic speech recognition with a Pytorch backend\footnote{https://github.com/espnet/espnet/tree/master/egs/commonvoice/asr1}. The base conformer model contains $12$ encoder layers, and $6$ decoder layers, with $0.1$ dropout rate. The encoder and decoder layers contain $2048$ hidden units. The input layer is a $2$D convolutional layer with a kernel size of $15$. Macaron implementation is used as described in \citep{conformer,espnetconformer}. For attention, four $256$-dimensional self-attention heads are used with relative positional encodings, swish activation and $0$ dropout.  For training, batch size of $32$ is used, with gradient accumulation for $2$ steps and gradients are clipped to $5$. The initial learning rate for the NoAM optimizer is set to $1.0$. Each model is trained for $50$ epochs. Beam search is used during decoding with beam width $10$, without any penalties or length restrictions, and without invoking any external language model. Model averaging is performed on  weights from $10$ epochs with the best validation accuracy. Other implementation details are the same as in Section~\ref{sec:appasr}.

\subsubsection{Speech Translation: Libri-Trans}
We use the ESPnet speech translation framework~\citep{espnetst} for implementing end-to-end speech translation (ST) with a Pytorch backend\footnote{https://github.com/espnet/espnet/tree/master/egs/libri\_trans/st1}. The base model for ST consists of an ASR encoder and a machine translation (MT) decoder. We initialized the encoder with a SpecAugment-pretrained ASR Transformer and the decoder with a pretrained MT Transformer.

\paragraph{Pretrained ASR Encoder.} The pretrained ASR transformer's architecture consists of $12$ encoder layers and $6$ decoder layers with a dropout rate of $0.1$. The encoder and decoder layers contain $2048$ hidden units. For attention, four $256$-dimensional attention heads are utilized. The NoAM optimizer is used with $5.0$ initial learning rate, $25000$ warmup steps, gradient accumulation for $2$ steps and gradients clipped at $5$.  Label smoothing is also applied with weight $0.1$. The batch size is set to $64$ for training. The weights from the $5$ models with best validation accuracy are averaged and used for initialization in the ST model.

\paragraph{Pretrained MT Decoder.} The pretrained MT transformer's architecture consists of $6$ encoder layers and $6$ decoder layers with a dropout rate of $0.1$. The encoder and decoder layers contain $2048$ hidden units. For attention, four $256$-dimensional attention heads are utilized. The NoAM optimizer is used without patience with $1.0$ initial learning rate, $8000$ warmup steps and gradients clipped at $5$. The weights of the transformer are initialized using the Uniform Xavier initialization~\citep{xavier}. Label smoothing is also applied with weight $0.1$. The batch size is set to $96$ for training. After training, the weights from the best $5$ models according to validation accuracy are averaged and used for initialization in the ST model.

After initialization, finetuning is performed for $50$ epochs. The optimizer used is the NoAM optimizer with $2.5$ initial learning rate, no patience and $25000$ warmup steps. The gradients are clipped at $5$, with gradients accumulated for every $2$ steps. The dropout rate utilized is $0.1$. For evaluation, model averaging is performed on weights from the $5$ epochs with best validation accuracy. A beam size of $10$ is utilized for decoding. Evaluations are performed on the predefined dev and test splits mentioned in Libri-Trans.

\subsubsection{Sound Classification: ESC-50 and UrbanSound8K}
\label{sec:soundclassapp}
We use the CNN10 architecture~\citep{panns} that is pretrained on AudioSet and takes mel-spectrograms as input. We use the official implementation\footnote{https://github.com/multitel-ai/urban-sound-classification-and-comparison}, and follow the experimental setup used in~\citep{urban}. For training, the Rectified Adam optimizer is used with LookAhead with an initial learning rate of $0.001$, $k=5$ and $\alpha=0.5$, early stopping is implemented with patience set to $30$. The batch size is set to $64$. 
\looseness=-1
Standard evaluation protocols defined for these benchmarks are followed: For ESC-50, cross-validation is performed over the $5$ pre-defined folds, and for UrbanSound8k, cross-validation is performed over the $10$ pre-defined folds. We use $2$ masks for FM, with a maximum frequency width of $8$. The $\alpha$ parameter for mixup (MX) is set to $1$. For TM and \SpA, $N=2$ and $T=24$. 

\subsubsection{Music Classification: GTZAN}
We use the official implementation\footnote{https://github.com/qiuqiangkong/panns\_transfer\_to\_gtzan}, and follow the experimental setup in~\citep{panns} and use the CNN14 architecture, pretrained on AudioSet, which is then further fine-tuned on data from GTZAN. The model is trained for $1000$ iterations with a batch size of $16$. The Adam optimizer with AMSGrad is used following an initial learning rate of $0.0001$. For evaluation, $10$-fold cross validation is performed. The parameters used for FM and MX are the same as in Section~\ref{sec:soundclassapp}; for TM and \SpA, $N=2$ and $T=64$. 

\subsubsection{Representation Learning: CLAR}
ResNet18 is the base encoder model with the output dimension set to $512$. The projection head used to generate $\mathbf{z}$ is implemented as three fully-connected layers with ReLU activations. To evaluate the learned representations, a linear classifier is trained on the frozen features, as has been done in prior work~\citep{grill2020bootstrap,chen2020exploring,zbontar2021barlow,chen2020simple}. We use the official implementation\footnote{https://github.com/haideraltahan/CLAR}, and follow the same experimental setup as CLAR, i.e. we perform our experiments on the Speech Commands-10 dataset consisting of speech samples corresponding to $10$ isolated-word commands. The batch size is set to $512$, and global batch normalization is used.%
We found the SGD optimizer to be more stable during training, compared to the Adam optimizer used in CLAR. We use Layer-wise Adaptive Rate Scheduling (LARS), with a learning rate of $1$, weight decay $10^{-4}$, with linear warm-up for the first $10$ epochs; the learning rate is decayed with a cosine schedule without restarts.

\subsubsection{Confidence Intervals \& Significance Testing}
We adopt the bootstrapping-based confidence interval estimation proposed in ~\citep{bootstrap}. The core idea is to create replications of a statistic by repeated random sampling of the dataset with replacement. For each sentence $i$, we record the number of words predicted correctly ($C_i$), the number of substitution errors ($S_i$), the number of insertion errors ($I_i$) and number of deletions ($D_i$). Let $X = (C_1, S_1, I_1, D_1),...(C_n, S_n, I_n, D_n)$ denote the errors we found in the test set. Next, we generate a bootstrap sample (with replacement) by repeating the sampling process $B=10^3$ times, each time yielding $X^{*b}=(C_1^{*b}, S_1^{*b}, I_1^{*b}, D_1^{*b}),...(C_n^{*b}, S_n^{*b}, I_n^{*b}, D_n^{*b})$ for $b=1..B$. For each $b$, we now have an estimate $W^{*b}$ of the WER. The uncertainty is then quantified in terms of the standard error of the $B$ samples.

For significance testing, we perform the Matched Pairs Sentence-Segment Word Error Test (mapsswe). The code we use for performing this test is part of the standard NIST Scoring Toolkit (SCTK) that is installed as part of Kaldi~\cite{kaldi}. For the significance test, we use a 95\% confidence interval to reject the null hypothesis. 
\subsection{Assets}
\subsubsection{Datasets}
\paragraph{LibriSpeech.} \citep{librispeech} is a collection of 960 hours of audiobooks that are a part of the LibriVox project. Most of the audiobooks come from Project Gutenberg. The training data is split into 3 partitions of 100hr, 360hr, and 500hr sets while the dev and test data are split into `clean’ and `other’ categories, respectively, depending upon whether the ASR system should be tested against clean or more challenging speech samples. Each of the dev and test sets is around 5hr in audio length. The dataset is open-sourced under the CC BY 4.0 license.
\paragraph{CommonVoice.} \citep{commonvoice} is an open-source project released by Mozilla specifically for ASR. Data for the project is collected from volunteers via the CommonVoice website and suitably anonymized before release. 
The dataset is distributed under the public domain Creative Commons (CC0) license.
\paragraph{TIMIT.}  The TIMIT~\citep{garofolo1993timit} Acoustic-Phonetic Continuous Speech Corpus is a standard small-vocabulary benchmark used for the evaluation of ASR systems. It consists of recordings from 630 speakers of 8 dialects of American English, with each speaker reading 10 phonetically-rich sentences. It also comes with  word and phone-level transcriptions of the speech. Research-only licensing for this data is available. More details can be found at \url{https://www.ldc.upenn.edu/language-resources/data/obtaining}. 
\paragraph{LibriTrans.} \citep{libritrans} is a speech translation dataset constructed by aligning foreign-language e-books with English utterances of LibriSpeech. The speech recordings and source texts are originally from the Gutenberg Project, a digital library of public domain books contributed by volunteers. The dataset is covered under the Creative Commons Attribution 4.0 license.
\paragraph{SpeechCommands.} \citep{warden2018speech} has $105,829$ one-second long utterances of $30$ short words, from $2,618$ speakers, contributed by volunteers through the AIY website\footnote{https://aiyprojects.withgoogle.com/open\_speech\_recording}. It is released under a Creative Commons BY 4.0 license. Each audio file contains a recording of a single spoken English word from a limited vocabulary. The dataset contains 35 labels (words) such as one-digit numbers, action oriented words, and arbitrarily short words. CLAR~\citep{clar} uses a simpler version ($\approx$20K samples) with only the utterances of the one-digit numbers.
\paragraph{ESC-50.} \citep{piczak2015dataset} is a labeled collection of 2000 environmental audio recordings suitable for benchmarking methods of environmental sound classification. The dataset consists of 5-second long recordings organized into 50 semantic classes, with 40 examples per class. Clips in this dataset have been manually extracted from public field recordings gathered by the FreeSound project\footnote{https://freesound.org/}. The dataset has been prearranged into 5 folds for comparable cross-validation, making sure that fragments from the same original source file are contained in a single fold. The dataset is available under the terms of the Creative Commons Attribution Non-Commercial license. A smaller subset (clips tagged as ESC-10) is distributed under CC BY (Attribution).
\paragraph{UrbanSound8K.} \citep{Salamon:UrbanSound:ACMMM:14} is an audio dataset that contains 8732 labeled sound excerpts (<=4s) of urban sounds from 10 classes: air\_conditioner, car\_horn, children\_playing, dog\_bark, drilling, engine\_idling, gun\_shot, jackhammer, siren, and street\_music. The classes are drawn from the urban sound taxonomy. All excerpts are taken from field recordings uploaded to FreeSound. The dataset is offered free of charge for non-commercial use only under the terms of the Creative Commons Attribution Noncommercial License (by-nc), version 3.0.
\paragraph{FSDNoisy18K.} \citep{fsd} is a crowdsourced audio classification dataset created as part of Freesound Annotator. It contains $18,532$ audio clips across $20$ classes, totalling $42.5$ hours of audio. The clip durations range from $300$ms to $30$s. The dataset is divided into a training set containing $17,585$ clips and a test set containing $947$ clips. The test set has been manually verified. In contrast, only $10\%$ of the training set is verified. It has been estimated that $45\%$ of the unverified labels are incorrect, and that $84\%$ of the incorrect labels are Out-of-Distribution.
\paragraph{GTZAN.} \citep{gtzan} is the most-used public dataset for evaluation in machine listening research for music genre recognition (MGR). The files were collected in 2000-2001 from a variety of sources including personal CDs, radio, microphone recordings, in order to represent a variety of recording conditions. The dataset consists of 1000 audio tracks each 30 seconds long. It contains 10 genres, each represented by 100 tracks. The tracks are all 22050Hz Mono 16-bit audio files in .wav format. The dataset is from Marsyas (Music Analysis, Retrieval and Synthesis for Audio Signals) and is distributed under the GNU Public Licence (GPL) Version 2.
\subsubsection{Frameworks}
\paragraph{Pytorch} ~\citep{pytorch} is an optimized tensor library for deep learning with CPU and GPU support. The framework supports several inbuilt computations such as automatic differentiation and GPU acceleration. The framework is licensed by Facebook, and it is similar to a BSD license. 
\paragraph{Pytorch-Lightning} ~\citep{lightning} disentangles the model details from the engineering details, leading to a hardware-agnostic and reproducible pipeline. This open-source toolkit is covered under the Apache License, Version 2.0 and is patent-pending.
\paragraph{ESPNet.} \citep{espnet, espnetst} is an open-source end-to-end speech processing platform. It is designed for the quick development of Speech Recognition, Machine Translation, Text-to-Speech and Speech Translation systems. This open-source toolkit is covered under the Apache License, Version 2.0. 
\paragraph{SpeechBrain.} \citep{speechbrain} is an open-source and all-in-one speech toolkit based on PyTorch that can be used to develop state-of-the-art speech technologies, including systems for speech recognition, speaker recognition, speech enhancement and multi-microphone signal processing. This open-source toolkit is covered under the Apache License, Version 2.0.  
\paragraph{DeepSpectrumLite.} \citep{deepspectrumlite}  is a Python toolkit to design and train light-weight Deep Neural Networks (DNNs) for classification tasks from raw audio data. The trained models run on embedded devices. This open-source toolkit is released under the GNU Public Licence (GPL) Version 3
\paragraph{Kaldi.} \citep{kaldi} is a speech recognition toolkit especially developed for research in ASR. It provides several standard codes/recipes for feature preparation, model training, statistical significance, and forms a core dependency within the ESPnet framework. The software is open sourced under the Apache License version 2.0. 
\subsection{\SpA with Sample Adaptive Policies}
Several recent works have investigated learning policies for data augmentation~\citep{autoaugment,randaugment}, and these policies have shown to help greatly in a wide range of tasks. Among these works, RandAugment~\citep{randaugment} is one of the most widely used method, which improves upon AutoAugment~\citep{autoaugment} by reducing the search space for policy search. However, both RandAugment and AutoAugment learn constant policies. SapAugment is a recently proposed framework to learn sample-adaptive policies for data augmentation, i.e. each training data point is augmented differently based on its training loss. A policy $f$ in SapAugment, parameterized by hyper-parameters $s, a$ is defined as 
$$
f_{s, a}(l)=1-I\left(s(1-a), s \cdot a ; l_{\text {rank }} / B\right)
$$
where $l_{\text {rank }}$ is the rank of loss value $l$ in the mini-batch, $B$ is the mini-batch size, and $I(\alpha, \beta ; x)$ represents the incomplete beta function. The output of $f$ determines the amount of augmentation applied. For $N$ augmentations, $N$ such policies, i.e $f_{s_1, a_1}, ..., f_{s_N, a_N}$ are learned jointly with their selection probabilities, $p_1, ..., p_N$. 

We follow~\citep{deepspectrumlite} which implements learned SapAugment  policies for Audio Classification. We perform our experiment on ESC-10 dataset, and similar to~\citep{deepspectrumlite}, we apply a linear classifier on top of an ImageNet-pretrained \textsc{DenseNet}121~\citep{huang2017densely} base convolutional encoder.

\begin{table}[h]
    \centering
    \caption{Comparison of TM and \SpA, when used in conjunction with CutMix, as data augmentations in Sample Adaptive Policies for Audio Classification}
    \begin{tabular}{lr}
    \toprule
    Augmentation & Accuracy\\
    \midrule
    None & $87.5$\\
    CutMix + TM~\citep{deepspectrumlite}&  $88$\\
    CutMix + \SpA & $\mathbf{89.25}$\\
    \bottomrule
    \end{tabular}
    \label{tab:sapaug}
\end{table}

Table \ref{tab:sapaug} compares sample adaptive policies on Time-Masking and \SpA, when used with CutMix. For evaluation, $5$-fold cross-validation is performed on the predefined splits. We notice that policies with \SpA perform at par with policies using Time-Masking. 
\subsection{\SpA and Noisy Data}

\subsubsection{Contrastive Learning}
~\citep{ucl} propose to learn sound representations using the pretext task of contrasting views primarily computed via mixing of training examples with background noise, and data augmentations including Random Resized Crop (RRC), Frequency and Time Masking. Representations are learned using the NT-Xent Loss with a base ResNet-18 encoder. For both, unsupervised learning of representations and supervised evaluation of learned representations, the input to the model is $96$-band log-mel spectrograms. The SGD Optimizer is used with a momentum of $0.9$, and weight decay of $10^{-4}$. For contrastive learning, model is trained for $500$ epochs with a batch size of $128$, with an initial learning rate of $0.03$, and lowered by a factor of $10$ in epochs $325$, and $425$. For supervised evaluation, models are trained for $200$ epochs with an initial learning rate of $0.01$. 

\begin{table}[h]
    \centering
    \caption{Evaluation of representations obtained using contrastive learning on FSD18KNoisy dataset, comparing TM and \SpA as data transforms}
    \begin{tabular}{lrrrr}
    \toprule
    Augmentation & KNN Accuracy & Linear Probe & FineTune-Clean & FineTune-Noisy\\
    \midrule
         FM + TM~\citep{ucl,specaugment} &  $66.99$ & $69.00$ & $73.70$ & $69.45$\\
         FM + SO & $\mathbf{68.95}$ & $\mathbf{71.78}$ & $\mathbf{75.15}$ & $\mathbf{71.86}$\\ 
    \midrule
      RRC +  FM + TM~\citep{ucl} &  $70.26$ & $74.40$ & $\mathbf{75.08}$ & $72.47$\\
      RRC +  FM + SO & $\mathbf{71.16}$ & $\mathbf{75.79}$ & $\mathbf{75.08}$ & $\mathbf{73.50}$\\ 
    \bottomrule
    \end{tabular}
    \label{tab:ucl}
\end{table}

Three standard ways of evaluating representations are followed, $k$-Nearest Neighbour (KNN) evaluation, linear probes, and end-to-end fine-tuning. For KNN evaluation, each audio patch is assigned a prediction using majority voting on $k=200$ neighbouring labels, based on its encoder representation. Linear probe involves training a linear classifier on top of the frozen pre-trained representations learnt by the encoder, and end-to-end fine-tuning involves fine-tuning the model on a downstream classification task. We use the official implementation\footnote{https://github.com/edufonseca/uclser20}, thereby following the same evaluation setup as~\citep{ucl}. 

Table \ref{tab:ucl} compares TM and \SpA as transforms to generate augmented views for contrastive learning, in conjunction with FM and RRC. We note that \SpA consistently outperforms TM across all evaluation methods.  

\subsubsection{Pseudo Labeling}
\citep{ood} first train an auxiliary classifier on a small subset of data consisting of clean examples. This classifier is used to detect Out-of-Distribution (OOD) samples to relabel. Samples on which the classifier outputs a label with high confidence, that does not match the ground truth, are selected for relabeling. Such a label misprediction with high confidence  suggests the ground truth label might be noisy and/or the true label might be similar to one of the target classes. A mixup like operation is performed on the noisy label and the label assigned by the auxiliary classifier, and the resulting target is used to relabel the sample. A final classifier is then trained on the union of the clean dataset and the relabeled dataset. 

\begin{table}[h]
    \centering
    \caption{Test evaluation on the FSD18KNoisy dataset comparing TM and \SpA, used in conjunction with MX and FM, as data augmentations for the auxiliary classifier and final classifiers~\citep{ood}.}
    \begin{tabular}{lrr}
    \toprule
     Augmentation & AP & Accuracy\\
     \midrule
    MX + FM + TM~\citep{ood} & $87.3{\scriptstyle \pm 0.3}$ & $80.9{\scriptstyle \pm 0.3}$\\
    MX + FM + \SpA & $87.1{\scriptstyle \pm 0.3}$ & $80.8{\scriptstyle \pm 0.2}$\\
    \bottomrule
    \end{tabular}
    \label{tab:pseudolbl}
\end{table}

We use the official implementation\footnote{https://github.com/tqbl/ood\_audio}, and follow the same experimental setup as~\citep{ood}. An ImageNet-pretrained DenseNet-201~\citep{huang2017densely} was used as the base classifier. The classifier was trained for $40$ epochs with a batch size of $128$, using an Adam optimizer~\citep{adam} with an initial learning rate of $0.0005$ decayed by $10\%$ every $2$ epochs. Table~\ref{tab:pseudolbl} compares TM and \SpA as augmentations for both the auxiliary and the final classifier, showing \SpA  performs comparably to TM.

\subsection{Wav2vec2 Fine-tuning}
Wav2vec2~\citep{wav2vec2} is a state of the art framework for self supervised learning of speech representations. The Wav2vec2 model consists of a convolutional feature encoder, which is applied on the raw waveform to obtain latent encodings, and a transformer, which subsequently uses these latent encodings as input and produces contextual representations as output. For training, latent representations are masked and a contrastive objective is optimized over quantized speech
representations. \citep{wav2vec2} show that Wav2vec2, pre-trained on $53$k hours of unlabeled speech, performs at par with previous methods while using orders of magnitude less labeled data. 

We compare TM and \SpA as data augmentations for fine-tuning a pre-trained Wav2Vec2 model on the TIMIT dataset for ASR. We follow the recipe\footnote{https://github.com/speechbrain/speechbrain/tree/develop/recipes/TIMIT/ASR/seq2seq} for the same task released by SpeechBrain~\citep{speechbrain}. Along with TM and FM, the recipe also uses speed pertubation. 

\begin{table}[h]
    \centering
    \caption{Finetuning Wav2vec2 on TIMIT for ASR using different data augmentation methods}
    \begin{tabular}{lrr}
    \toprule
         Augmentation & Dev PER & Test PER\\
         \midrule
         FM + TM~\citep{speechbrain} & $7.21$ & $8.54$\\
         FM + \SpA & $7.34$ & $8.59$\\
         FM + TM + \SpA & $7.31$ & $8.58$\\
    \bottomrule
    \end{tabular}
    \label{tab:timit}
\end{table}

Table \ref{tab:timit} evaluates the Phone Error Rate (PER) obtained when fine-tuning with data augmentation methods, TM and \SpA, on the dev and test splits. We notice that \SpA achieves comparable performance with TM, validating the applicability of \SpA on \textbf{raw waveforms}. 

\subsection{\SpA for Representation Learning with Raw Waveforms}

A similar setup as described in Section 4.3 is used, except ResNet18 operations such as convolutions, max pooling and batch normalization are switched from 2D to 1D
because the model takes raw audio signal as input. Table~\ref{tab:clarraw} compares TM with \SpA when used in conjunction with FD for both semi-supervised and self-supervised representation learning. \SpA yields consistent performance improvements compared to TM, especially in the self-supervised setting.
\looseness=-1

\begin{table}[h]
    \centering
    \caption{Comparing classification accuracies using \SpA and TM in semi-supervised and self-supervised settings on the Speech Commands Dataset.}
    \begin{tabular}{ll|r}
    \toprule
         Type & Method & Accuracy\\
         \midrule
         Semi-Supervised & CLAR (FD + TM)~\citep{clar} & $94.8$\\
         &CLAR (FD + \SpA) & $\mathbf{94.9}$\\
         \midrule
         Self-Supervised & SimCLR (FD + TM)~\citep{clar} & $71.0$\\
         & SimCLR (FD + \SpA) & $\mathbf{77.4}$\\
    \bottomrule
    \end{tabular}
    \label{tab:clarraw}
\end{table}

\subsection{ASR for Multiple Languages: CommonVoice}
In addition to the experiments in Section 4.1.2, we train ASR models for six different languages spanning training sizes ranging from $27$ hours to $116$ hours: Estonian, Czech, Dutch, Portugese, Esperanto and Russian. We use a conformer architecture adapted for ESPnet~\citep{espnetconformer}. Experimental setup is similar to Section 4.1.2, and implementation details can be found in Section \ref{sec:appcommon}. 

\begin{table}[h]
    \centering
    \caption{Evaluation of TM and \SpA, when used with TW and FM, using test WERs across multiple different languages.}
    \begin{tabular}{lrrrrrrrr}
    \toprule
     Augmentation  & Estonian  &  Czech & Dutch & Portugese & Esperanto & Russian\\
      & 27 hrs & 29 hrs & 45 hrs & 53 hrs & 89 hrs & 116 hrs\\ 
    \midrule
    TW + FM + TM~\citep{specaugment,espnet}& $41.2$  & $16.3$ & $2.1$ & $10.5$ & $\mathbf{14.1}$& $9.4$\\
    TW + FM + \SpA & $\mathbf{40.8}$ & $\mathbf{15.7}$ & $\mathbf{2.0}$ & $\mathbf{10.1}$ & $14.4$ & 
      $\mathbf{9.0}$\\
    \bottomrule
    \end{tabular}
    \label{tab:commonvoiceapp}
\end{table}

Table~\ref{tab:commonvoiceapp} shows WERs for six languages. Our experiments demonstrate that \SpA offers comparable performance, and often outperforms TM.

\subsection{\SpA and Frame Skipping}
\label{app:frameskipping}
We compare \SpA with Frame Skipping/Dropping methods~\cite{dynamicframeskipping,45555,7472084,6639137} and list a few reasons why \SpA might be advantageous in comparison
\begin{itemize}
    \item Flexibility in model output: By design, frame skipping methods impose the constraint that all frames in the skip interval should have the same state/output. In contrast, \SpA does not enforce any such constraint making it more flexible.
    \item Data and parameter efficient: Stochasticity in \SpA achieves the best of both worlds: The number of possible distinct samples per input is high and scales with the length of the input thus resulting in higher amounts of training data compared to static frame skipping, and \SpA avoids training a separate model to calculate skip intervals like in dynamic frame skipping methods.  
    \item Easier integration with existing methods: \SpA has been proposed as a drop-in replacement of Time Masking, and as such can be integrated into implementations of existing systems for speech and audio tasks with minimal changes. The use of \SpA in conjunction with common augmentation techniques like Mixup, Frequency Masking, etc. is also straightforward. We believe integrating frame skipping would not be as seamless. 
    \item Performance gains: Static and dynamic frame skipping make computations much faster, but only lead to marginal improvements in performance (if at all); cf., Table 1 in \cite{dynamicframeskipping}. In contrast, \SpA (like Time Masking in SpecAugment) improves generalization and yields significant improvements on test set evaluations.
    \item Better Internal LM for ASR: By completely deleting information within an interval, \SpA defers to the internal LM to fill in the gaps, thereby (implicitly) facilitating learning a better language model. 
    \item Applications outside ASR: Most works using frame skipping/dropping methods have done so in the context of ASR. \SpA is shown to be effective in several settings, including audio classification, representation learning etc, and it is unclear if frame skipping methods would help in such settings without further experimentation. 
\end{itemize}

\bibliographystyle{unsrtnat}
{\small \bibliography{references}}

\end{document}